\documentclass[apsrev4,pra,notitlepage]{revtex4-1} 
\usepackage{graphicx}
\usepackage{amssymb,calc}
\usepackage{amsmath,amsfonts,amssymb}
\usepackage{graphicx}
\usepackage[hidelinks]{hyperref}
\hypersetup{
 colorlinks=false,
 pdfauthor={Martin Fally@univie},
 pdftitle={Diffraction theories for off-Bragg replay: J.T. Sheridan's seminal work and consequences},
 pdfkeywords={Holography, Diffraction theory, John T. Sheridan}
 }
\usepackage[version=4]{mhchem}

\begin{document}
\title{Diffraction theories for off-Bragg replay: J.T. Sheridan's seminal work and consequences}
\author{Martin Fally}
\affiliation{University of Vienna, Faculty of Physics, Boltzmanngasse 5, A-1090 Wien, Austria}
\email{martin.fally@univie.ac.at}
\begin{abstract}
Based on the seminal work by \textit{John T. Sheridan} \,\cite{Sheridan-jmo92} we discuss the usefulness and validity of simple diffraction theories frequently used to determine and characterize optical holographic gratings. Experimental investigations obtained in recent years highlight the correctness of his analysis which favours an alternative approach over the most widely used \textit{Kogelnik} theory.
\end{abstract}
\keywords{Holography, Diffraction theory, John T. Sheridan}
\maketitle
\section{Introduction}\label{I1}

In the early 1990ies \textit{J.T. Sheridan} published a paper comparing three different analytic theories and a numerical approach of diffraction from one-dimensional volume holographic gratings for significantly off-Bragg re\-con\-struction \,\cite{Sheridan-jmo92}. He argued that the prevalent first-order theory by \textit{H. Kogelnik} \,\cite{Kogelnik-Bell69} is inferior to the less well known first-order theory by \textit{N. Uchida} \,\cite{Uchida-josa73} in the off-Bragg case. His results are vindicated by a comparison to a second-order coupled-mode theory by \textit{J.A. Kong} \,\cite{Kong-josa77} and the rigorous coupled-wave theory as introduced by \textit{Moharam, Gaylord} and coworkers \,\cite{Moharam-josa81,Gaylord-apb82,Moharam-josaa95}. The results are important when it comes to the direction of propagation of the diffracted wave and its phase \,\cite{Sheppard-ije76}. While the outcome of his paper was disputed at the time of publication, recent detailed experimental investigations confirmed his conclusion and highlighted the correctness of his analysis \cite{Fally-apb12,Prijatelj-pra13}.

In this contribution we will discuss the aforementioned first-order theories, their differences and the applicability to experimental data. The aim is to provide a background as well as a concise hitchhiker's guide to data evaluation of one-dimensional holographic gratings that might be useful for non-specialists, e.g., engineers, chemists and those who make use of holographic gratings (astronomers, spectroscopists) in science and industry. We will thus elaborate on the way of characterization of holographic gratings by using diffraction experiments and the evaluation of relevant parameters from experimental data. We focus in particular on the angular dependence of the diffraction efficiency. A vast number of advanced books beyond the simple assumptions taken here are recommended to the interested reader \,\cite{Loewen-97,Popov-14,Bao-22}. 

\section{Diffraction from holographic gratings}
\subsection{The grating}
\begin{figure}[h]\centering
\includegraphics[width=\textwidth]{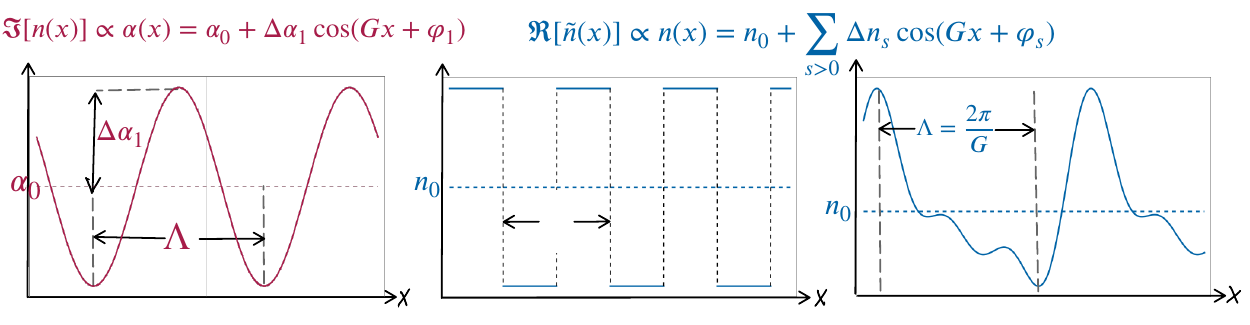}
\caption{\label{fig:gratings} Examples for grating profiles. \textit{(left)} Sinusoidal, pure absorption grating, Eq. (\ref{eq:alpha}) with $\Delta\alpha_s\equiv 0$ for $s\geq 2$. \textit{(center)} Binary, pure refractive-index grating, Eq. (\ref{eq:refin}) with $\Delta n_s\neq 0$ and $\varphi_s=-\pi/2$. \textit{(right)} Pure refractive-index grating with general profile (asymmetric), Eq. (\ref{eq:refin}) with $\Delta n_s\neq 0$ and $\varphi_s\neq 0,\pi/2,\pi$.}
\end{figure}

We consider the particular case of holographic gratings which are recorded by the interference of two plane waves with wavelength $\lambda$ (and same linear polarization state, say $s$-polarization) at a recording angle $2\Theta_B$. This results in a sinusoidal, one-dimensional light pattern with a period $\Lambda=\lambda/(2\sin\Theta_B)$ along the direction $x$ (without loss of generality) which is perpendicular to the bisectrix of the beams. When exposing a light sensitive material to this interference pattern, a periodic structure is generated therein.
Generally, such one-dimensional periodic structures for light in a medium are described by either their permittivity or their refractive index which might be a complex quantity:
\begin{equation}
 \label{eq:refinx}
 \tilde{n}(x)=\sum_{s=-\infty}^{\infty}\tilde{n}_s e^{\imath s G x}, \quad \tilde{n}_s\in\mathbb{C}.
\end{equation}
Here, $G=2\pi/\Lambda$ is the spatial frequency 
and $|\tilde{n}_s|=n_s\in\mathbb{R}$ the amplitude of the Fourier component for the $s$-th order. We distinguish two different limiting cases: (1) pure phase gratings and (2) pure absorption gratings. In the first case we will end up with a real-valued refractive index 
\begin{equation}
\label{eq:refin}
n(x)=n_0+\sum_{s=1}^{+\infty}\underbrace{2|\tilde{n}_s|}_{\Delta n_s}\cos(s Gx+\varphi_s),\in\mathbb{R} \quad\text{for}\quad\tilde{n}_{-s}=\tilde{n}_s^*\in\mathbb{C}
 \end{equation}
with $n_0$ the average refractive index of the medium and $\Delta n_s$ the refractive index modulation of order $s$.
The second case results in an imaginary-valued refractive index which is related to the absorption constant $\Im{[\tilde{n}(x)]}=\alpha(x)\lambda/(4\pi)$ by:
\begin{equation}\label{eq:alpha}
 \alpha(x)=\alpha_0+\sum_{s=1}^{+\infty}\underbrace{2|\tilde{n}_s|}_{\Delta \alpha_s}\cos(s Gx+\varphi_s), \in\mathbb{R}; \quad\text{for}\quad\tilde{n}_{-s}=-\tilde{n}_s^*\in\mathbb{C}.
\end{equation}
$\alpha_0$ is the mean absorption constant and $\Delta\alpha_s$ is the absorption modulation of order $s$.
For the general case of mixed gratings including a phase shift $\phi$ between them we refer to Refs.\,\cite{Guibelalde-oqe84,Sutter-josab90,Kahmann-josaa93,Carretero-ol01,Fally-oex08,Ellabban-oex06}.
These Fourier cosine-coefficients, the relative phases $\varphi_s$ and the grating period $\Lambda$ determine -- together with the thickness $d_0$ and the employed (reconstruction) wavelength $\lambda$ -- the grating's diffraction properties. 
Fig.\,\ref{fig:gratings} shows a schematic of various holographic gratings illustrating important variants of the general profile Eq. (\ref{eq:refinx}).

A typical problem is to find out the characteristics of a particular, given grating, which are:
\begin{enumerate}
\item the type of the grating (absorption grating, phase grating or mixed grating, respectively)
\item the grating profile with its Fourier coefficients (e.g. sinusoidal, binary\ldots)
\item the symmetry of the grating with its phases (e.g., rectangular, blazed, sawtooth\ldots).
\item the grating thickness $d_0$.
\end{enumerate}
To start with the investiagtion we introduce and consider an important quantity, which can be obtained experimentally as well a theoretically: \textbf{the diffraction efficiency}. From an experimental point of view it  is given by the ratio of the diffracted light intensity into order $m$ and the incident light intensity 
\begin{equation}
\label{eq:DEexp}
\eta_{m,\text{exp}}=\frac{I_m}{I_\text{in}}.
\end{equation}
For pure lossless phase gratings Eq.\,(\ref{eq:DEexp}) is equivalent to the diffracted intensity into order $m$ over the sum of all diffracted orders, an experimentally quite useful definition:
\begin{equation}
 \label{eq:DEexp_phase}
\eta_{m,\text{exp}}=\frac{I_m}{\sum\limits_{m}^{}I_{m}}.
\end{equation}
For pure phase gratings with an average absorption Eq.\,(\ref{eq:DEexp}) can be written as (see Sec.\,\ref{sec:grating_type}):
\begin{equation}
 \label{eq:DEexp_phase}
\eta_{m,\text{exp}}=e^{-\alpha_0d}\frac{I_m}{\sum\limits_mI_{m}}.
\end{equation}

\subsection{The diffraction geometry}
\begin{figure}[h]\centering
\includegraphics[width=\textwidth-1cm]{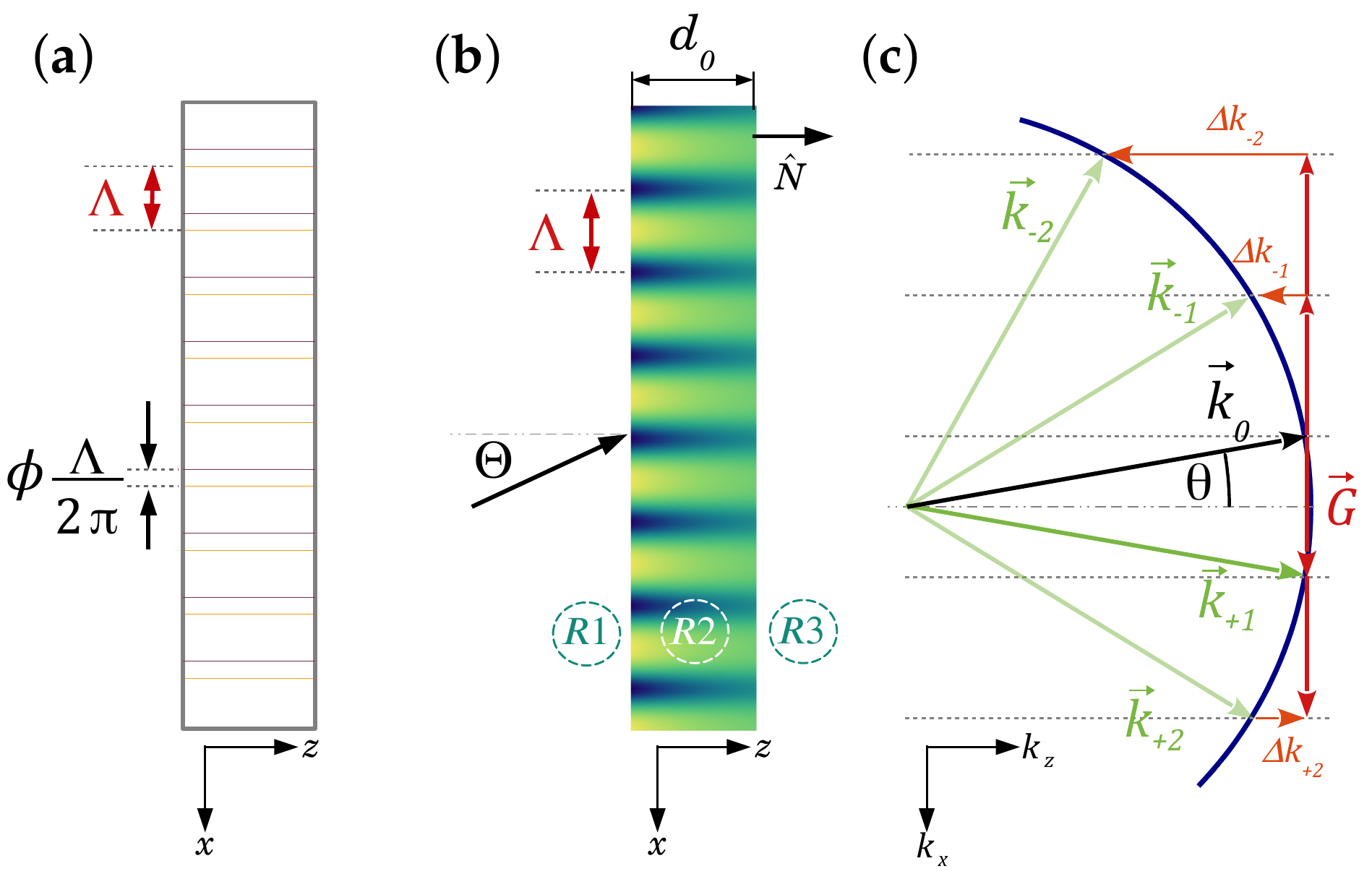}
\caption{\label{fig:geometry} Diffraction geometry for transmission gratings, see Ref.\,\,\cite{Ellabban-Materials17}. (a) sketch for mixed gratings with phase shift $\phi$ between absorption grating and refractive-index grating. (b) grating with arbitrary profile (and a decay along the $z$-direction, neglected in this paper). (c) diffracted wavevectors $\vec k_m$ in reciprocal space. Further not yet discussed notation will be given in Sec.\,\ref{sec:model}.}
\end{figure}

Here we specialize on planar unslanted transmission gratings which are among the most important geometries used. In Fig.\,\ref{fig:geometry} the grating geometry as well as the diffraction geometry for the wavevectors is shown. The term unslanted refers to the fact that grating vector and sample surface normal are mutually perpendicular, $\vec G\perp\hat N$. For further discussion we divide the space in three regions: the entrance region (R1), the grating region (R2), and the exit region (R3). Such unslanted transmission gratings deliver the major amount of intensities in the forward direction into region R3. In what follows we assume that the refractive index in R1 and R3 is homogeneous and $n_0$, thus no reflection effects from boundaries need to be considered.
The typical experimental data are obtained by measuring Bragg-detuning curves of the diffraction efficiency $\eta_{m,\text{exp}}$ either by rotating the grating by an angle $\theta$ around an axis $y$ which is perpendicular to $\vec G\times\hat N$ or by scanning the wavelength. Both methods allow for determining the relevant quantities $\tilde{n}_s,\varphi_s$ under investigation. Standard diffraction experiments, however, only allow to extract information on the modulus $|\tilde{n}_s|$, the phase-information is lost. Advanced experimental techniques or evaluation schemes are required to retrieve also these values \cite{Fally-oex21}.

\section{Modelling the diffraction}\label{sec:model}
Modelling the diffraction of quanta (x-rays,  electrons) from periodic structures has been the subject of study for more than a century. First theories came up with the discovery of x-rays and their distinct diffraction features from perfectly periodic single crystals \,\cite{Darwin-PhilMag14a,Darwin-PhilMag14b,Ewald-adp16.1,Ewald-adp16.2,Ewald-adp17.1,Ewald-adp17.2,Bethe-adp28,Bloch-zfp28}. The solutions to the wave equation - arising from either Maxwell's equations or the Schrödinger equation - for a periodic potential (a complex dielectric permittivity) were found. Those Bloch-waves which are consistent with energy and momentum conservation, i.e., obeying Bragg's law, impose certain conditions on the diffracted wavevectors. Typically only \textbf{two modes} are considered: the forward diffracted wave (zero order) and another one (plus or minus first order, depending on the combination of wavelength and angle of incidence). We call this the \textbf{modal approach} for two waves according to Ref.\,\,\cite{Gaylord-apb82}. It is characteristic that each of the Bloch-waves is an eigensolution to the wave equation and propagates through the grating without change.

\begin{figure}[h]
\begin{minipage}{\textwidth}
 \includegraphics[width=\textwidth/2-1mm,trim=0 0 0 5,clip]{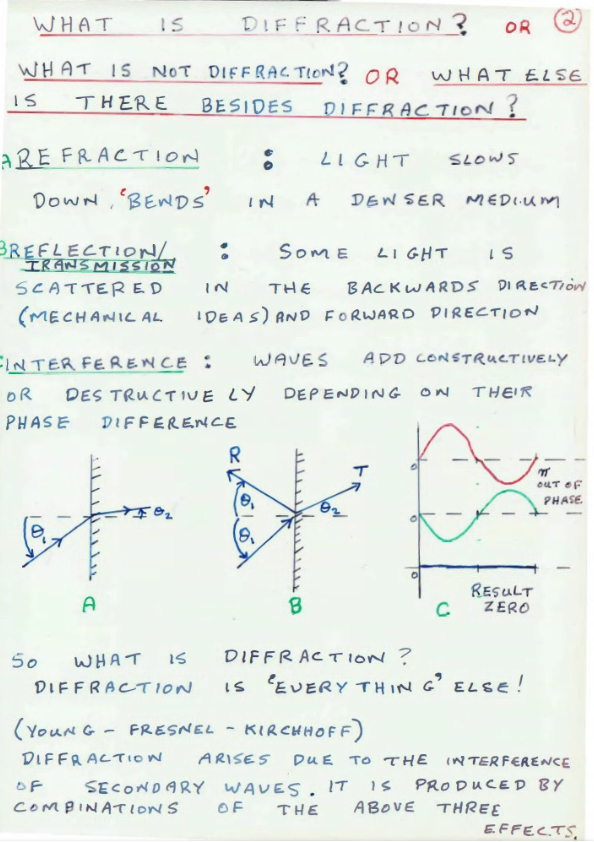}
 \includegraphics[width=\textwidth/2-1mm,trim=5 0 10 10,clip]{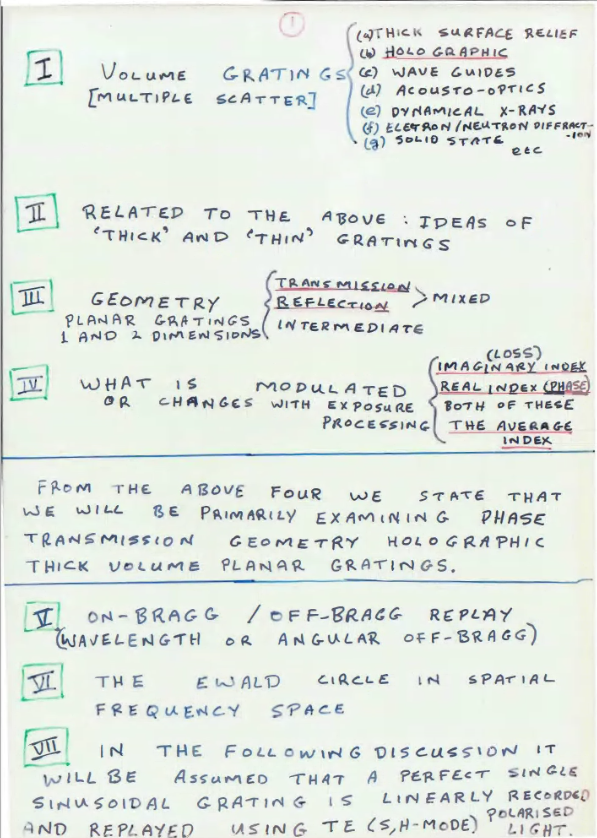}
\end{minipage}
\caption{\label{fig:sean_slides} Introductory slides of \textit{J.T. Sheridan} during his plenary video-talk on the occasion of the 50$^\text{th}$ anniversary of Dennis Gabor’s Nobel Prize Award (May 20, 2021) entitled: ``Holography and Engineering the Future'' by John T. Sheridan (University College Dublin, Ireland)}
\end{figure}

However, there is an alternative equivalent, in some sense more practical, description for the problem of diffraction from a grating: the coupled wave theory. Here, \textit{J.T. Sheridan} contributed in his early career \,\cite{Sheridan-Optik90b,Sheridan-Optik90a,Sheridan-Optik90} and ever since then, in particular also by educating young as well as advanced researchers when introducing the intricacies of diffraction theories to them \,\cite{Sheridan-jmo92,Sheridan-josaa92,Sheridan-oc92,Sheridan-josaa93,Sheridan-Optik93,Sheridan-Optik94,Sheridan-oc94,Sheridan-josaa94}. Related original slides of one of his plenary talks on the occasion of commemorating the 50$^\text{th}$ anniversary of Dennis Gabor’s Nobel Prize Award -- organized by the University of Alicante, Spain, in May 2021 -- during the pandemic are shown in Fig.\,\ref{fig:sean_slides}.
In what follows we will discuss several important approaches to the diffraction problem and point out \textit{J.T. Sheridan's} insight.
\subsection{The Raman-Nath approach}
In the 1930-ies diffraction of light from periodic standing acoustic waves was examined experimentally by \textit{Debye} and \textit{Sears} and explained in detail in a series of papers by \textit{Raman} and \textit{Nath}. They employed an approach based on the phase-changes of light occuring when propagating through a periodic (sinusoidal) medium. They first assume normal incidence of light, neglecting any dephasing due to oblique incidence \,\cite{Raman-piasa36}. In the subsequent paper they include oblique incidence and arrive at an equation for maximum diffraction at certain angles $\theta_m$ which actually are defined by the grating equation \,\cite{Raman.2-piasa36}:
\begin{equation}
 \label{eq:gratingequation}
 \sin(\theta_m)-\sin(\theta_0)=m\frac{\lambda}{\Lambda}, m\in\mathbb{Z}.
\end{equation}
Here, $\theta_0$ is the angle of the incident beam's wavevector $\vec k_{in}$ with sample surface normal $\hat N$ (as well as that of the zero order diffracted wavevector), and $\theta_m$ is the angle between $\hat N$ and the $m$-th order diffracted wavevector $\vec k_{m}$ (see Fig.\,\ref{fig:geometry}). This is an application of the Floquet theorem Eq. (\ref{eq:Floquetcondition}) discussed later. By considering path-length differences (phases) of light beams, \textit{Raman} and \textit{Nath} finally were able to deliver a simple and famous formula for the diffraction efficiency (relative intensities) of the $m$-th diffraction order $\eta_m$:
\begin{equation}
  \label{eq:DE_RN}
\eta_m(\theta)=J^2_m\left[\frac{2\nu}{\cos(\theta)}\text{sinc}\left(\frac{\pi d \tan(\theta)}{2\Lambda}\right)\right].
\end{equation}
for each of these orders. Here, $J_m$ is the Bessel function of first kind and $m$-th order, $\theta$ the angle of incidence and $\nu=n_1\pi d/\lambda$ the grating strength for a pure phase grating. Of course, there exists just a single Fourier component $n_1$ for a sinusoidal grating. For the original problem $\Lambda\gg\lambda$ and $d\approx\Lambda$ (\textit{``bending of light rays within the medium may be ignored provided the total depth of the cell is not excessive''}\,\cite{Raman.2-piasa36}) we approximate $\sin\theta\approx\tan\theta\approx\theta,\cos\theta\approx 1$ and the angular dependence of the diffraction efficiency (nearly) vanishes with a diffraction efficiency for the $m$-th order of 
\begin{equation}
\eta_m\approx J_m^2(2\nu),\quad \Lambda\gg\lambda, d\approx\Lambda.
\end{equation}
This is a frequently used result when dephasing (angular dependence) can be neglected. It describes the diffracted power into various orders at the same time, i.e., diffraction occurs in a particular regime which then was coined \textbf{Raman-Nath diffraction regime} as will be discussed in Sec.\,\ref{sec:diffraction_regime}. A similar result can be obtained by employing the concept of amplitude transmittance functions in Fourier optics \,\cite{Goodman-05}.

\subsection{The coupled wave theory 1: Kogelnik's approach for two waves}
While this approach was applicable for quite a number of diverse diffraction problems from periodic media, it was not useful for example in x-ray diffraction from perfect single crystals. The general reason is that implicitely the first Born approximation (undepleted pump beam, no multiple scattering) is not valid for thick perfect crystals. The crystallographers code-word for diffraction in thick perfect crystals was said to be described by a ``dynamical theory'' rather than a ``kinematical theory''. As already stated above corresponding theories were developed in the early twentieth century, nicely summarized for x-rays in Ref.\,\cite{Zachariasen-45}, the classical book by \textit{Zachariasen}, or in a more recent article \,\cite{Batterman-rmp64}. Interestingly enough, for light optics the relevance of this field became apparent only after the discovery of the laser and the increasing importance of holographic techniques. In the late sixties of the last century the off-axis holographic technique \,\cite{Leith-josa62} and its application in thick holographic recording media \,\cite{Chen-apl68} asked for a theory of diffraction from artificially created,  one-dimensionally periodic, planar volume holograms (gratings). After a number of papers were highlighting various particular limited cases \,\cite{Klein-jasa65,Burckhardt-josa66,Burckhardt-josa67,Gabor-prsa68}, a major step forward was done by \textit{H. Kogelnik} in his celebrated and extensively cited article \,\cite{Kogelnik-Bell69}. He gives a detailed derivation, based on a paper by \textit{Phariseau}\,\,\cite{Phariseau-piasa56}, of the relevant equations for the diffraction power (diffraction efficiency) from volume holograms for
\textbf{both}, transmission holograms and reflection holograms in the following important cases:

\begin{enumerate}\addtolength{\itemsep}{-0.35ex}
\item lossless dielectric gratings
\item lossy dielectric gratings
\item unslanted absorption gratings
\item slanted absorption gratings
\item mixed gratings
\end{enumerate}

The interesting part of his extensive theory is the alternative approach to the ``dynamical theory of diffraction'' which is called \textbf{coupled wave theory}. He limits his consideration to the case of only \textbf{two waves} propagating at the same time in the \textbf{sinusoidal} grating whenever the Bragg condition is (almost) fulfilled. These two waves are coupled via the periodic potential (grating) with a certain coupling constant $\kappa$. The latter gives rise to an exchange of energy between the amplitudes $S_0(z),S_1(z)$ (which he denotes by $R,S$ remembering the terms \textbf{r}eference and \textbf{s}ignal beam from holography) of the waves during propagation. The theory boils down to solve a coupled system of linear differential equations for these amplitudes. For phase gratings the coupling constant is a real quantity, $\kappa=\Delta n_1\pi/\lambda\in\mathbb{R}$, and for absorption gratings an imaginary number, $\kappa=\imath\Delta\alpha_1\pi/\lambda\in\mathbb{C}\backslash \mathbb{R}$, mixed gratings finally have $\kappa\in\mathbb{C}$. 

In the frame of this approach the Helmholtz scalar-equation $[\vec\nabla^2+(2\pi n(x)/\lambda)^2]E(x,z)=0$ is solved employing the Bragg condition, Eq. (\ref{eq:Braggcondition}), and an \textit{Ansatz} for $E(x,z)=S_0(z)\exp{(\imath \vec k_0\cdot\vec x)}+S_1(z)\exp{(\imath \vec k_1\cdot\vec x)}$ in the grating region R2. This results in a system of two coupled, homogeneous, linear differential equations of second order in $S_0(z), S_1(z)$. To facilitate the solution  of this task, Kogelnik disregards second order derivatives (slowly varying envelope approximation). This is justified whenever the coupling strength $\nu=\kappa d/\cos\theta$ is not ``too large'', which can be assumed valid in most of the relevant cases. However, this simplification also bears \textbf{another, more important, consequence} which is related to the wavevector $\vec k_1$, i.e. the \textbf{direction of the diffracted beam}. Actually, the simplfication results in a reduction of the number of boundary conditions from four to two. Thus the remaining boundary conditions lead to (physically) ambiguous results. Or in other words: a manifold of solutions exists which gives mathematically correct results, however, they might not be meaningful in physics. This problem of Kogelnik's solution was clearly stated already in the publications by \textit{Syms} and \textit{Solymar}\,\cite{Syms-ao83,Syms-oa85,Syms-90}.

Kogelnik's choice for the diffracted wavevector $\vec k_1$ is simply
\begin{equation}
\label{eq:KogWV}
\vec k_1=\vec k_0+\vec G
\end{equation}
and is termed ``K-vector-closure-method'' (KVCM). The resulting diffraction efficiency of the first order (for an unslanted, lossless transmission grating) then reads:
\begin{eqnarray}
\label{eq:DE_Kog}
\eta_1(\theta)&=&\nu^2\text{sinc}^2\left(\sqrt{\nu^2+\xi^2}\right),\\
\nu&=&\frac{n_1\pi d}{\lambda\cos(\theta)},\\
\xi&=&\frac{\pi d}{\Lambda\cos(\theta)}\left(\sin(\theta)-\frac{\lambda}{2n_0\Lambda}\right)\approx\frac{\pi d}{\Lambda}(\theta-\theta_1).
\end{eqnarray}
Here, $\theta_1=\arcsin[\lambda/(2n_0\Lambda)]$ is the Bragg angle (in the medium) and in the last step we have linearized the Bragg-angle detuning. 

\subsection{The coupled wave theory 2: Uchida's approach for two waves}
An alternative choice for the diffracted wavevector was suggested by \textit{N. Uchida} \,\cite{Uchida-josa73}, In his publication he focusses on providing a theory and resulting equations for the diffraction efficiency for an \textbf{attenuated grating} with an exponential decay of the modulation $\Delta n_1(z)=\Delta n_1(z=0)\exp(-z/L)$ along the sample depth. This is definitely a very valuable contribution and the work is continuously cited for this reason. However, unfortunately his physically more favourable choice of the wavevector is hidden behind the former, in particular as the author himself remarks that {\em ``the present treatment for the wave-vector mismatch is somewhat different from that by Kogelnik [\ldots] However, no essential difference arises in the result of calculation.''}\,\cite{Uchida-josa73} Definitely, this is bare understatement as far as (1) the direction of the diffracted beam and the (2) off-Bragg diffraction efficiency are concerned.

Uchida's choice for the diffracted wavevector accounts for phase matching at the exit boundary, energy- and momentum conservation and is sometimes called the ``Beta-value-method'' (BVM) in literature:
\begin{equation}
 \label{eq:UchWV}
 \vec k_1=\vec k_0+\vec G+\Delta k_1\hat N,
\end{equation}
where $\Delta k_1$ is found by the condition $|\vec k_1|=|\vec k_0|=2\pi n_0/\lambda=\beta$. Taking into account the geometry shown in Fig.\,\ref{fig:geometry} we find
\begin{eqnarray}\nonumber
\Delta k_1&=&-k_{0,z}\pm\sqrt{k_{0,z}^2-G(2k_{0,x}+G)},\quad G=2\pi/\Lambda. \\
&=&-\beta\left[\cos(\theta)-\sqrt{1-(\sin(\theta)-G/\beta)^2}\right],\quad\theta>0.
\end{eqnarray}
The geometry for Uchida's choice is shown in Fig.\,\ref{fig:geometry} (and on-Bragg for the first diffraction order, i.e., $\Delta k_{+1}=0$). The diffraction efficiency then, with redefined parameters $\nu$ and $\xi$, reads:
\begin{eqnarray}
 \label{eq:DE_Uch}
\eta_1(\theta)&=&\frac{c_R}{c_S}\nu^2\text{sinc}^2\left(\sqrt{\nu^2+\xi^2}\right),\\
c_R&=&\cos(\theta)\\
c_S&=&\sqrt{1-(\sin(\theta)-G/\beta)^2},\quad\theta>0\\
\nu&=&\frac{n_1\pi d}{\lambda\sqrt{c_Rc_S}},\\
\xi&=&\beta(c_R-c_S)d/2.
\end{eqnarray}

\subsection{The coupled wave theory 3: general solution by Gaylord and Moharam}\label{sec:RCWA}
At this stage \textit{J.T. Sheridan} decided to compare the two simple first-order theories \,\cite{Kogelnik-Bell69,Uchida-josa73} to the second-order theory \,\cite{Kong-josa77} and the full and rigorous solution (rigorous coupled wave analysis, RCWA) \,\cite{Moharam-josa81}. \textit{Gaylord} and \textit{Moharam} treated the diffraction problem rigorously by solving the Helmholtz equation in each of the regions, matching the corresponding solutions at the entrance as well as at the exit boundaries for both, the electric and the magnetic field components. The total field in the grating region R2 is expanded in an infinite Fourier series of waves with amplitudes $S_m(z)$:
\begin{equation}
\label{eq:RCWA}
E(x,z)=\sum_{m=-\infty}^{+\infty}S_m(z)e^{-\imath \vec k_m\cdot\vec x}.
\end{equation}
Therefore, this method is also called the Fourier modal method \,\cite{Bao-22}. Note, that in contrast to the modal theory a single wave is not a solution to the Helmholtz-equation. The resulting system of an infinite number of coupled linear differential equations of second order can be transformed into a state space description which allows to find the solutions by seeking for eigenvalues and eigenvectors of a characteristic matrix. Unknown coefficients are uniquely determined by the boundary conditions. Naturally, the diffraction efficiency cannot be written in simple analytical terms, though. We just refer to a number of excellent publications for a large diversity of cases \,\cite{Moharam-josa81,Gaylord-apb82,Moharam-josa82,Moharam-josa82a,Moharam-josaa86,Moharam-josaa95,Moharam-josaa95.02,Lalanne-josaa96,Bao-22}.

By utilizing this general solution, the parameters relevant for the different diffraction regimes could be quantified (for a sinusoidal grating)\,\cite{Gaylord-ao81,Moharam_b-oc80,Moharam_rn-oc80}, the validity of approximate theories, their hierarchy and relation to the modal approach\,\cite{Gaylord-apb82} could be explored.

\begin{figure}[h]
\includegraphics[width=\textwidth]{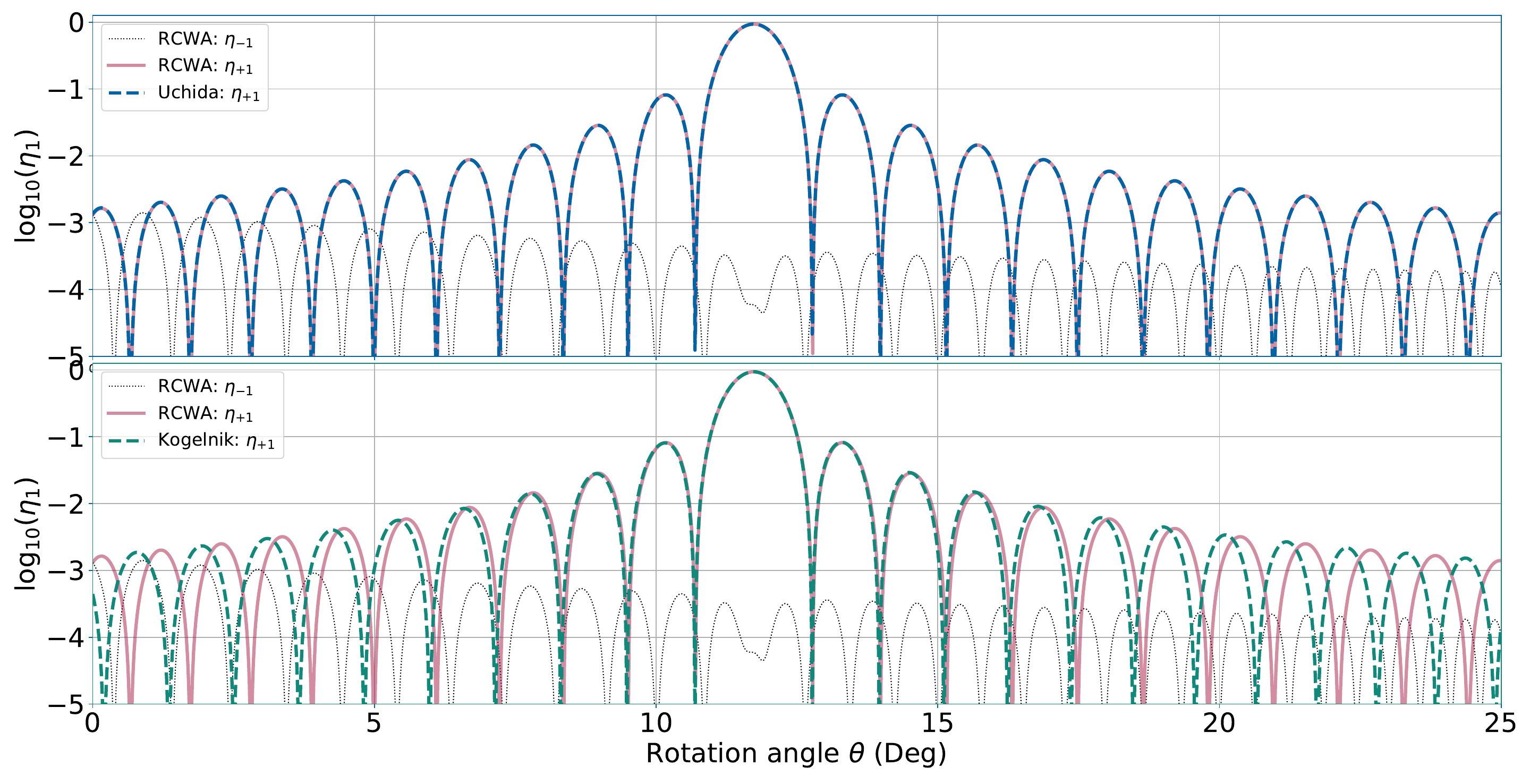}
\caption{\label{fig:Comp_RCWA} Angular dependence of the diffraction efficiency $\eta_1(\theta)$ in the Bragg-regime (logarithmic plot). Comparison between the rigorous approach (3-waves-RCWA, red) and the approximate first-order theories. \textit{(top)} Uchida's theory using BVM (blue) and \textit{(bottom)} Kogelnik's theory using KVCM (mint). Parameters: $d=40~\mu$m, $\Lambda=0.8~\mu$m, $\lambda=0.488~\mu$m, $\Delta n_1=5\times 10^{-3}$. Dotted black lines are the $\eta_{-1}$ from RCWA. Further information see text.}
\end{figure}
As a consequence \textit{Sheridan} thus compared, from a theoretical point of view, both of the first-order theories with the second-order theory and the rigorous approach. In a diligent analysis he came to the conclusion that \textbf{Uchida's choice of wavevectors is superior} to that of Kogelnik's \,\cite{Sheridan-jmo92,Sheridan-josaa94}. 
In Fig.\,\ref{fig:Comp_RCWA} the diffraction efficiency $\eta_1(\theta)$ for the RCWA and Uchida's first-order approach (top panel) and Kogelnik's first-order approach (bottom panel) are compared (for a sinusoidal grating), respectively. It can be clearly seen that Uchida's result (blue) cannot be distinguished from the 3-wave-RCWA (red), i.e. $m=0,\pm 1$, even in the far off-Bragg region, whereas Kogelnik's prediction (mint) becomes increasingly dephased for off-Bragg angles. For convenience the minus first diffraction order in the 3-wave-RCWA is depicted as dotted black lines, too. Even when neglecting this order could play a role in the vicinity of normal incidence ($\theta=0$) this will not be the case for ($\theta\gg\theta_1$ for which the KVCM is considerably mismatched, the BVM however remains indistinguishable from the 3-wave-RCWA. This basic theoretical result underlines the validity of \textit{J.T. Sheridan's} analysis who obtained it for the first time.

\section{Experimental observations: diffraction regimes and types of gratings}
From a pragmatic point of view one typically is interested to extract the decisive parameters ($\Delta n_s,\Delta \alpha_s,$ $\varphi_s,\phi,d_0,\Lambda$) from experimental data by employing least-squares fitting routines for $\eta_{m,\text{exp}}(\theta)$ using the easiest one of the above diffraction theories that allows to model the data. The classification and characterization of the experimental data will be the topic of the following subsections.
\subsection{Diffraction regimes}\label{sec:diffraction_regime}
\begin{figure}[h]
\includegraphics[width=\textwidth]{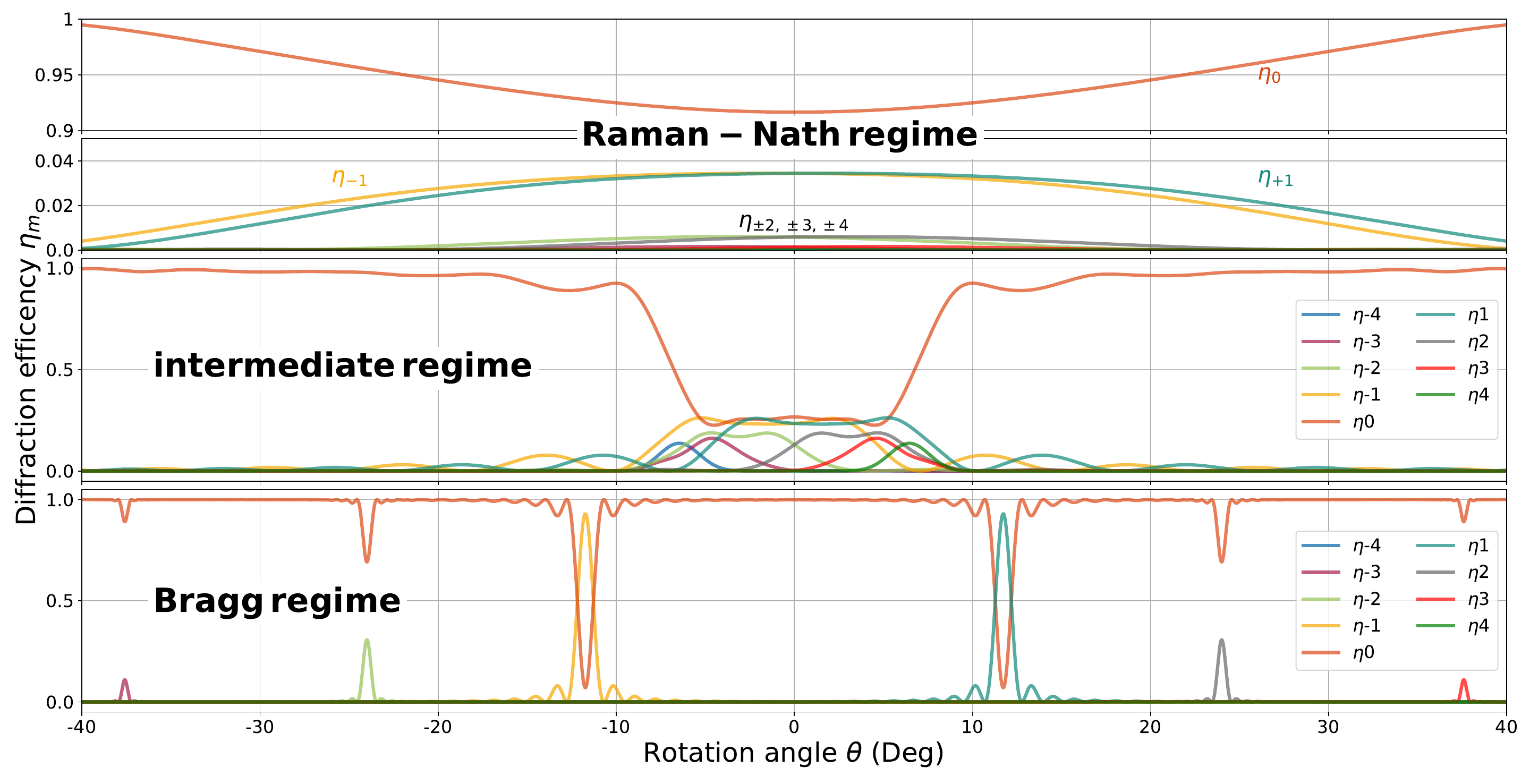}
\caption{\label{fig:diffreg} Angular dependence of $\eta_m(\theta)$ using the (realistic) simulation parameters $\lambda=488\,$nm, $\Delta n_{1,2,3}=(5,2,1)\times 10^{-3},\varphi_{1,2,3}=(0,\pi,0)$. \textit{(top two panels)} \textbf{Raman-Nath} diffraction regime ($\Lambda=6\,\mu$m, $d_0=6\,\mu$m); \textit{(central panel)} \textbf{intermediate} diffraction regime ($\Lambda=6\,\mu$m, $d_0=40\,\mu$m);  \textit{(bottom panel)} \textbf{Bragg} diffraction regime ($\Lambda=0.8\,\mu$m, $d_0=40\,\mu$m).} 
\end{figure}
Let us start our discussion with the type of diffraction regime as theoretically discussed in Sec.\,\ref{sec:RCWA}. We can identify three distinct regimes, at least for sinusoidal gratings, for which diffraction occurs \,\cite{Gaylord-ao81}:
\begin{enumerate}
\item the Raman-Nath diffraction regime \,\cite{Gaylord-ao81,Moharam_rn-oc80}.  In this case \textbf{many} waves propagate in the grating \textbf{at the same time} along directions (orders) with a variety of wavevectors $k_m$ having low but considerable amplitudes, see Fig.\,\ref{fig:diffreg} top panel. These permitted wavevectors $k_m$ for the waves which travel along diffraction directions at angles $\theta_m$ are given by the Floquet-condition for periodic potentials:
\begin{equation}\label{eq:Floquetcondition}
 \vec k_m=\vec k_0+m\vec G,
\end{equation}
where $\vec k_0$ is the wavevector of the incident (and forward diffracted) beam and $\vec k_m$ that of the $m$th-order diffracted beam; for a grating which is periodic only along the $x-$direction, relation (\ref{eq:Floquetcondition}) is required only for the $k_x-$components \,\cite{Gaylord-apb82}. This is also called the ``thin'' grating regime. 
\item the Bragg diffraction regime. Here, only \textbf{two} diffraction orders with reasonable amplitudes propagate \textbf{at the same time}, the $0$-th and another one ($m$-th order). This is sometimes also called the ``thick'' grating regime \,\cite{Gaylord-ao81,Moharam_b-oc80} for which the diffraction efficiency is noticeable only for the zeroth and the $m$-th order which obeys (at least approximately) the Bragg condition:
\begin{equation}\label{eq:Braggcondition}
 2\Lambda n_0\sin(\theta_m)=m\lambda.
\end{equation}
In other words: for a certain range of angles in the vicinity of the corresponding Bragg condition with $m=+1$ the $+1$-st order is strongly diffracted; in another range of angles say the $-3$-rd order might be diffracted without noticeable intensity to other directions (except for the zeroth), see Fig.\,\ref{fig:diffreg} bottom panel.
\item the \textbf{intermediate regime} for which neither of the cases is true.
\end{enumerate}
(Realistic) simulation data illustrating the different diffraction regimes assuming a pure phase grating are shown in Fig.\,\ref{fig:diffreg}.
Note, that a non-sinusoidal phase grating with three Cosine-Fourier-coefficients $\Delta n_{1,2,3}$ and relative phases $\varphi_{1,2,3}=(0,\pi,0)$ was assumed. Therefore, $\eta_{+m}(\theta_{+m})=\eta_{-m}(\theta_{-m})$.
\subsection{Grating types}\label{sec:grating_type}
To discriminate between the different grating types for a sinusoidal grating from an experimental point of view, one takes a measured data set, $I_m(\theta)$, and will be able to distinguish the following cases:
\begin{itemize}
 \item pure phase gratings, or
 \item pure absorption gratings, or
 \item mixed gratings (with/without additional phase shift $\phi$)
\end{itemize}
\begin{figure}[h]
\includegraphics[width=\textwidth-2mm]{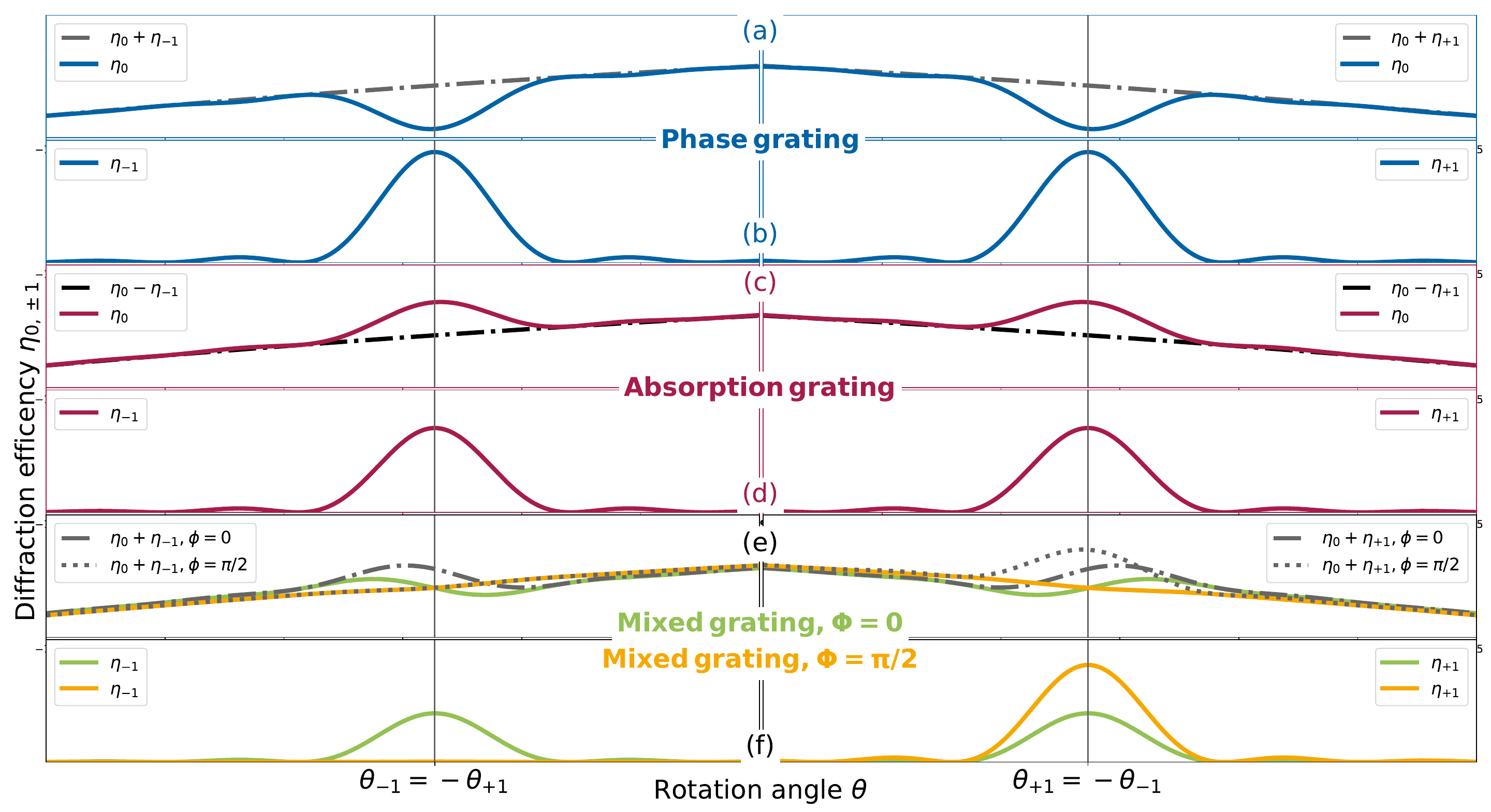}
\caption{\label{fig:gtypes} Angular dependence of the diffraction efficiencies of $\eta_{0,\pm 1}$ for a sinusoidal grating in the vicinity of the corresponding Bragg angles $\theta_{-1}$ (\textit{left}) and $\theta_{+1}$ (\textit{right}). The top two panels \textit{(a),(b)} show the behaviour for \textbf{pure phase gratings}, the two central panels \textit{(c),(d)} the situation for \textbf{pure absorption gratings}, and the two lower panels \textit{(e),(f)} for \textbf{mixed gratings}. The latter cases depend on the relative phase-shift $\phi$ between the phase grating and the absorption grating. Gray lines correspond to the sum of all diffraction orders, black lines to $\eta_0-\sum_{m\neq 0}\eta_m$. Reflections from surfaces are neglected by phase-matching R1-R2-R3, i.e., assuming the same $n_0$ for these regions.} 
\end{figure}
These cases are illustrated for Bragg diffraction and a sinusoidal grating in Fig.\,\ref{fig:gtypes}.
Note, that $\eta_{+1}(\theta_{+1})\neq\eta_{-1}(\theta_{-1})$ can also occur for non-sinusoidal pure phase gratings whenever $\varphi_s\neq r\pi,r\in\mathbb{Z}$. However, for the sum of diffracted intensities $\sum_m I_m(\theta)= e^{-\alpha_0d_0/\cos(\theta)}$ will always hold in the latter case of pure phase gratings.

\subsection{Check- and to-do-lists for practicioners}
Classification and evaluation of light optical diffraction data from a one dimensional grating requires a number of steps which are concisely presented below.

The first step, of course, is to obtain the experimental data required: the angular (or sometimes wavelength) dependence of the intensities $I_m(\theta)$ diffracted from the grating under investigation. 
\subsubsection*{Diffraction regime check}
To discriminate between the Bragg- and the Raman-Nath diffraction regimes, respectively, is easily done by shining a laser beam at perpendicular incidence on the grating and placing a screen a few ten centimeters apart. If more than the transmitted (or better: the forward diffracted) beam can be seen, diffraction takes place in the Raman-Nath regime. Detectors can be placed at positions of high intensities, i.e. at angles given by the grating equation Eq. (\ref{eq:gratingequation}) for further, quantitative characterization.

On the contrary, if no more than the a single beam is observed on the screen, a rotation around an axis $y$ perpendicular to the grating vector and the sample surface normal by angles $\theta$ has to be performed. It is expected that at certain angles $\theta_m$, given by the Bragg condition Eq. (\ref{eq:Braggcondition}), bright spots appear on the screen and the forward diffracted, zero order beam is attenuated (or enhanced). Thus a first measurement of the intensity of the latter will give direct evidence for the related Bragg-angles $\theta_m$.

A final decision on the diffraction regimes, in particular the intermediate regime, asks for a full quantitative analysis: $\eta_m(\theta)$ must be measured. 

\subsubsection*{Type of grating check}
Typically photodiodes with detection areas of a few mm$^2$ in combination with amplifiers are used to ensure a wide dynamic range of several orders of magnitude in optical power. Therefore, the detectors have to be placed at angles $\theta_m$ at which diffraction occurs. While this sounds trivial one has to consider that diffraction angles $\theta$ do not only depend on the grating period $\Lambda$ and the wavelength $\lambda$ but also on its detuning from Eq. (\ref{eq:Braggcondition}). Note that \textbf{only if the Bragg condition} is fulfilled the corresponding diffraction angle $\theta=\theta_m=\arcsin(m\lambda/(2n_0\Lambda))$, which forms also \textbf{the minimum angle} $2\theta_m$ between the zero order and the $m$-th order diffracted beam. For off-Bragg situations the enclosed angles become larger with an excess angle for both deviations $\theta<\theta_m$ as well as $\theta>\theta_m$ \cite{Fally-apb12}. From a practical point of view this might lead to a beam which moves out of the detection area and repositioning must be considered.

To start with finding out the grating type from experimental data, we define fingerprints for each of these cases is as follows:
\begin{itemize}
 \item pure phase gratings (with absorption loss but without absorption grating) for which energy is conserved in the diffraction process. They are characterized by the fact that the \textbf{sum} of diffracted intensities of the zero-order $I_0$ and all other orders $I_m$ yields a smooth function $\sum_m I_m(\theta)= e^{-\alpha_0d_0/\cos(\theta)}$; moreover $\eta_{+1}(\theta_{+1})=\eta_{-1}(\theta_{-1})$ holds where $\theta_{\pm 1}$ is the Bragg angle for the $\pm 1$-st diffraction order, see Eq. \ref{eq:Braggcondition}
 \item pure absorption gratings for which energy is dissipated in the diffraction process. They are characterized by the fact that the \textbf{difference} of diffracted intensities of the zero-order $I_0$ and the sum of all other orders $I_m, m\neq 0$ yields a smooth function $I_0-\sum_{m\neq 0} I_m(\theta)= e^{-\alpha_0d_0/\cos(\theta)}$; again $\eta_{+1}(\theta_{+1})=\eta_{-1}(\theta_{-1})$ holds. Here, in the vicinity of the Bragg angles losses of the diffracted intensity are \textbf{reduced} as compared to the average absorption (``Borrmann effect'').
 \item for mixed gratings the behaviour depends on the relative values of the real and imaginary parts of $\Re{[\tilde{n}_{\pm 1}]},\Im{[\tilde{n}_{\pm 1}]}$ and the relative phase shift $\phi$ between them. In general $\eta_{+1}(\theta_{+1})\neq\eta_{-1}(\theta_{-1})$ (except for $\phi=0,\pi$). For details see Ref.\,\,\cite{Carretero-ol01} and in particular Fig.\,2 of  Ref.\,\,\cite{Fally-oex08}.
\end{itemize}
\subsubsection*{Type of theory to use}
The advice for the appropriate theory that fits the deduced experimental $\eta_{m,\text{exp}}(\theta)$ and at the same time is the most simple theory, can be kept short. 

For \textbf{diffraction in the Bragg regime}, the most frequently employed configuration, with moderate uniform coupling constants $\kappa$ and assuming sinusoidal gratings the first-order theories \cite{Kogelnik-Bell69,Uchida-josa73} are sufficient to evaluate the data using Eqs.\,(\ref{eq:DE_Uch}) or (\ref{eq:DE_Kog}) reprinted here. This is true for pure phase gratings as well as pure absorption gratings. Whenever an (exponential) attenuation of the coupling constant with the grating depth comes into play \textit{Uchida's} publication delivers analytic solutions \cite{Uchida-josa73}. When it comes to mixed gratings including a phase shift between the phase and the absorption grating again analytic theories were published with slightly different focus, sometimes making use of interferometric (beam-coupling) techniques at Bragg incidence or by the angular dependence of $\eta_1(\theta)$ \cite{Guibelalde-oqe84,Sutter-josab90,Kahmann-josaa93,Carretero-ol01,Fally-oex08,Syms-ao83,Russell-ap80,Montemezzani-pre97,Neipp-jpd02,Neipp-oex02,Ellabban-spie07,Flauger-polymers19}. For high coupling constants, also called overcoupled or overmodulated gratings, a careful analysis requires second-order or rigorous theories \cite{Neipp-jmo04}: in a series of experiments together with a comparison to theories, the limitations of the first-order approaches were investigated by the \textit{Alicante} group. In particular the characteristics of the angular dependence of the diffraction efficiency for strong coupling, i.e. overmodulated gratings, was in the focus \,\cite{Neipp-joa01,Neipp-ao01,Neipp-jpd02,Neipp-oex02,Gallego-oc03,Neipp-Optik03,Neipp-oex03,Ortuno-ao03,Neipp-jmo04,Neipp-oc04,Neipp-Optik04,Gallego-oex05,Hernandez-oc06,Gallego-Materials12}.

For \textbf{diffraction in the Raman-Nath regime} and sinusoidal gratings the results were given in the corresponding publications \cite{Raman-piasa36,Raman.2-piasa36}, reprinted as Eq. (\ref{eq:DE_RN}) here. 

For more complicated, still ``thin'' gratings, rigorous theories are required. This is even more important for the \textbf{intermediate regime} and a big number of cases which were not addressed here, such as: nonsinusoidal gratings, binary and blazed gratings, diffraction considering the state of polarization (TE and TM modes), conical out-of-plane diffraction,\ldots. Still this can be mastered using analytic theories and taking into account a large number of Fourier-coefficients of Eq. (\ref{eq:refin}), also including the phase-shifts $\varphi_s$ (RCWA). However, when the periodic symmetry starts to be broken (defect structures), numerical solutions employing solvers are required. 
\section{Discussion and Conclusion}
After all the journey through the holographic grating landscape let us finally come to the announced comparison between the two first-order theories. Based on all findings above our goal is to make a definite decision on whether one of them is superior to the other. 
\subsection{Comparison between Kogelnik's and Uchida's approach: theoretical and experimental results}
\subsubsection*{Direction of the diffracted wavevector}
One rather obvious distinction between the two approaches which can be accessed experimentally is the direction of the diffracted beam. This was recognized early \cite{Sheppard-ije76} but never tested. Experiments on a volume holographic nanoparticle-polymer composite grating under rotations (in-plane) and conical diffraction clearly demonstrated that the direction of the beam follows the predictions given by \textit{Uchida}. In his approach the minimum angle between zero and first order diffracted beam is twice the Bragg-angle for the standard in-plane configuration. As explained above, any deviation will lead to an increase of this enclosed angle. In contrast, the KVCM choice would lead to an increase for angles $\theta>\theta_1$ and a decrease for $\theta<\theta_1$. The latter is not observed at all. This therefore is a very \textbf{direct experimental consequence} and proof of the correct choice of wavevectors according to the BVM. 

\subsubsection*{The off-Bragg detuning curve $\eta_1(\theta)$}
In the above comparison between the RCWA and the two first-order theories we have already found that the off-Bragg detuning curve for the BVM is indistinguishable from that of the RCWA, whereas the KVCM becomes worse for increasing deviation. For $\theta=\theta_1$  Eqs. (\ref{eq:DE_Kog}) and (\ref{eq:DE_Uch}) lead to the same result and are correct. However, for $\theta\neq\theta_1$ and the KVCM conceptual problems arise: the momentum-energy conservation is violated, a feature that can be nicely shown using the Ewald-sphere construction. Or pointed out more dramatically: the diffracted beam would have changed its wavelength! Definitely, this pseudo-inelastic scattering is inacceptable and becomes worse the further the detuning parameter $\xi$ deviates from zero. This issue may be of little relevance for very thick gratings, for example in mm-thick photorefractive electrooptic crystals with moderate coupling constants ($\kappa\leq 10^{-4}/\mu\text{m}$), where the diffracted intensity vanishes rapidly at small (angular or wavelength) detuning. Maybe this is one of the reasons why the community did not care too much about such unphysical solution. However, in contrary the problem is strongly intensified for say optimized photopolymers of high coupling constants ($\kappa\leq 10^{-1}/\mu\text{m}$) and low thicknesses in the $\mu$m range. Or as stated in Ref.\,\,\cite{Syms-ao83}: \textit{``Any differences between the two choices will be seen in a volume hologram when there is appreciable diffraction for large deviation from the Bragg condition, namely, in the sidelobe structure for comparatively thin holograms.''} Nowadays, excellent photopolymeric materials exist with saturated refractive-index modulations up to $\Delta n_1=0.04$ leading to coupling constants $\kappa=1/4/\mu\text{m}$, therefore, mechanical thicknesses of a few $10~\mu$m lead already to overcoupling with a $\eta(\theta_1)$ in the second quadrant. 

As an addition to the theoretical proof for the superiority of the BVM \cite{Sheridan-jmo92}, experiments were performed using a SiO$_\text{2}$-nanoparticle-polymer composite with a moderate thickness of about $d_0\approx 60~\mu$m and a resulting diffraction efficiency well in the first quadrant. It was shown that the experimental $\eta_{1,\text{exp}}(\theta)$, in particular the minima of this Bragg angle detuning curve, can be fitted using a rigorous or the first-order theory based on the BVM excellently. The KVCM on the other hand largely fails to reproduce the position of the minima, thus leading to large residuals \cite{Prijatelj-pra13}. This experimental result confirms the theoretical one and demonstrates that the phase of the diffracted beam is incorrect with the KVCM wavevector choice.
\subsection{Conclusion, summary and recommendation}
While it is obvious that over the last 50+ years the optics and holographic communities have favoured \textit{Kogelnik's} KVCM-approach to describe diffraction from volume holographic gratings, it is evident that the \textbf{equally simple} approach using the BVM delivers \textbf{much better results}. With this statement we mean that \textit{Uchida's} first-order theory:
\begin{enumerate}
 \item predicts the \textbf{direction of the diffracted beam correctly} which is confirmed by experiments\,\cite{Fally-apb12}
 \item shows the same \textbf{angular response} of the diffraction efficiency $\eta_1(\theta)$ as the experimentally obtained one\,\cite{Prijatelj-pra13},
 \item and \textbf{matches} the Bragg-angle detuning curve derived from a \textbf{rigorous coupled-wave theory }(RCWA) as already recognized by \textit{JT. Sheridan}\,\cite{Sheridan-jmo92} as well as that derived from the modal theory (dynamical theory of diffraction)\,\cite{Prijatelj-pra13}.
\end{enumerate}
Actually, \textit{J.T. Sheridan} let me know to have been enthusiastic about the work of \textit{Uchida: ``I have always been a strong admirer and promoter of Uchida’s work''}. He emphasized the subtle, however important, differences between the theories but still was realistic enough to note that \textit{``there are no Nobel prizes here and the community  – as you note - to have opted for the KVCM''} (e-mail by JTS in 2012). 

We therefore can and should conclude that one of the first-order theories, within its known limits, is \textbf{superior} to another, very well received approach. Thus we strongly recommend to \textbf{make use of \textit{Uchida's} approach} in future.

\section*{Acknowledgements}

The research described in this publication was made possible in part by the Austrian Science Fund (FWF) [P-35597-N] and the Austrian Research Promotion Agency (FFG), Quantum-Austria NextPi, grant number FO999896034.

The author would like to thank his long lasting collaborators J. Klepp, C. Pruner, M.A. Ellabban, I. Dreven\v{s}ek and Y. Tomita as well as his former students M. Prijatelj, G. Heuberger, G. Glavan and P. Flauger who performed part of the optical measurements to elucidate the described diffraction phenomena. The full set of original data is accessible on request in their theses.


\begin{thebibliography}{}
\providecommand{\url}[1]{\texttt{#1}}
\providecommand{\urlprefix}{~}
\addtolength{\itemsep}{1.2ex}
\bibitem{Sheridan-jmo92}
J.~T. Sheridan.
\newblock \textrm{A comparison of diffraction theories for off-{Bragg} replay}.
\newblock \emph{J. Mod. Optic} \textbf{39}, 1709 (1992).
\newblock \doi{10.1080/713823578}.

\bibitem{Kogelnik-Bell69}
H.~Kogelnik.
\newblock \textrm{Coupled wave theory for thick hologram gratings}.
\newblock \emph{AT\&T Tech. J.} \textbf{48}, 2909 (1969).
\newblock \doi{10.1002/j.1538-7305.1969.tb01198.x}.

\bibitem{Uchida-josa73}
N.~Uchida.
\newblock \textrm{Calculation of diffraction efficiency in hologram gratings
  attenuated along the direction perpendicular to the grating vector}.
\newblock \emph{J. Opt. Soc. Am.}  \textbf{63}, 280 (1973).
\newblock \doi{10.1364/JOSA.63.000280}.

\bibitem{Kong-josa77}
J.~A. Kong.
\newblock \textrm{Second-order coupled-mode equations for spatially periodic
  media}.
\newblock \emph{J. Opt. Soc. Am.}  \textbf{67}, 825 (1977).
\newblock \doi{10.1364/JOSA.67.000825}.

\bibitem{Moharam-josa81}
M.~G. Moharam and T.~K. Gaylord.
\newblock \textrm{Rigorous coupled-wave analysis of planar-grating diffraction}.
\newblock \emph{J. Opt. Soc. Am.}  \textbf{71}, 811 (1981).
\newblock \doi{10.1364/JOSA.71.000811}.

\bibitem{Gaylord-apb82}
T.~K. Gaylord and M.~G. Moharam.
\newblock \textrm{Planar dielectric grating diffraction theories}.
\newblock \emph{Appl. Phys. B}  \textbf{28}, 1 (1982).
\newblock \doi{10.1007/BF00693885}.

\bibitem{Moharam-josaa95}
M.~G. Moharam, E.~B. Grann, D.~A. Pommet, and T.~K. Gaylord.
\newblock \textrm{Formulation for stable and efficient implementation of the
  rigorous coupled-wave analysis of binary gratings}.
\newblock \emph{J. Opt. Soc. Am. A}  \textbf{12}, 1068 (1995).
\newblock \doi{10.1364/JOSAA.12.001068}.

\bibitem{Sheppard-ije76}
C.~J.~R. Sheppard.
\newblock \textrm{The application of the dynamical theory of x-ray diffraction to
  thick hologram gratings}
\newblock  \emph{Int. J. Electronics}  \textbf{41}, 365 (1976).
\newblock \doi{10.1080/00207217608920647}.

\bibitem{Fally-apb12}
M.~Fally, J.~Klepp, and Y.~Tomita.
\newblock \textrm{An experimental study on the validity of diffraction theories
  for {off-Bragg} replay of volume holographic gratings}.
\newblock \emph{Appl. Phys. B}  \textbf{108}, 89 (2012).
\newblock \doi{10.1007/s00340-012-5090-x}.

\bibitem{Prijatelj-pra13}
M.~Prijatelj, J.~Klepp, Y.~Tomita, and M.~Fally.
\newblock \textrm{{Far-off-Bragg} reconstruction of volume holographic gratings:
  A comparison of experiment and theories}.
\newblock \emph{Phys. Rev. A}  \textbf{87}, 063810:1 (2013).
\newblock \doi{10.1103/PhysRevA.87.063810}.

\bibitem{Loewen-97}
E.~G. Loewen and E.~Popov.
\newblock \textrm{Diffraction Gratings and Applications}.
\newblock Optical Science and Engineering. Taylor \& Francis, Boca Raton, 1st
  ed. (1997).
\newblock \doi{10.1201/9781315214849}.

\bibitem{Popov-14}
E.~Popov, ed.
\newblock \textrm{Gratings: Theory and Numeric Applications}.
\newblock ISBN: 978-2-85399-943-4. Presses Universitaires de Provence, second
  ed. (2014).
\newline\urlprefix\url{www.fresnel.fr/numerical-grating-book-2}

\bibitem{Bao-22}
G.~Bao and P.~Li.
\newblock \textrm{Maxwell’s Equations in Periodic Structures}, vol. 208 of
  \emph{Applied Mathematical Sciences}.
\newblock Springer \& Science Press Beijing (2022).
\newblock \doi{10.1007/978-981-16-0061-6}.

\bibitem{Guibelalde-oqe84}
E.~Guibelalde.
\newblock \textrm{Coupled wave analysis for out-of-phase mixed thick hologram
  gratings}.
\newblock \emph{Opt. Quant. Electron.}  \textbf{16}, 173 (1984).
\newblock \doi{10.1007/BF00620135}.

\bibitem{Sutter-josab90}
K.~Sutter and P.~G{\"u}nter.
\newblock \textrm{Photorefractive gratings in the organic crystal
  2-cyclooctylamino-5-nitropyridine doped with
  7,7,8,8-tetracyanoquinodimethane}.
\newblock \emph{J. Opt. Soc. Am. B}  \textbf{7}, 2274 (1990).
\newblock \doi{10.1364/JOSAB.7.002274}.

\bibitem{Kahmann-josaa93}
F.~Kahmann.
\newblock \textrm{Separate and simultaneous investigation of absorption gratings
  and refractive-index gratings by beam-coupling analysis}.
\newblock \emph{J. Opt. Soc. Am. A}  \textbf{10}, 1562 (1993).
\newblock \doi{10.1364/JOSAA.10.001562}.

\bibitem{Carretero-ol01}
L.~Carretero, R.~F. Madrigal, A.~Fimia, S.~Blaya, and A.~Bel{\'e}ndez.
\newblock \textrm{Study of angular responses of mixed amplitude-phase holographic
  gratings: shifted borrmann effect}.
\newblock \emph{Opt. Lett.}  \textbf{26}, 786 (2001).

\bibitem{Fally-oex08}
M.~Fally, M.~Ellabban, and I.~Dreven\v{s}ek-Olenik.
\newblock \textrm{Out-of-phase mixed holographic gratings : a quantative
  analysis}.
\newblock \emph{Opt. Express}  \textbf{16}, 6528 (2008).
\newblock \doi{10.1364/OE.16.006528}.

\bibitem{Ellabban-oex06}
M.~A. Ellabban, M.~Fally, R.~A. Rupp, and L.~Kov{\'a}cs.
\newblock \textrm{Light-induced phase and amplitude gratings in centrosymmetric
  gadolinium gallium garnet doped with {Calcium}}.
\newblock \emph{Opt. Express}  \textbf{14}, 593 (2006).
\newblock \doi{10.1364/OPEX.14.000593}.

\bibitem{Ellabban-Materials17}
M.~A. Ellabban, G.~Glavan, J.~Klepp, and M.~Fally.
\newblock \textrm{A comprehensive study of photorefractive properties in
  poly(ethylene glycol)dimethacrylate - ionic liquid composites}.
\newblock \emph{Materials}  \textbf{10}, 9 (2017).
\newblock \doi{http://dx.doi.org/10.3390/ma10010009}.

\bibitem{Fally-oex21}
M.~Fally, Y.~Tomita, A.~Fimia, R.~Madrigal, J.~Guo, J.~Kohlbrecher, and
  J.~Klepp.
\newblock \textrm{Experimental determination of nanocomposite grating structures
  by light- and neutron-diffraction in the multi-wave-coupling regime}.
\newblock \emph{Opt. Express}  \textbf{29}, 16153 (2021).
\newblock \doi{10.1364/OE.424233}.

\bibitem{Darwin-PhilMag14a}
C.~Darwin.
\newblock \textrm{{XXXIV.} the theory of {X-ray} reflexion}.
\newblock \emph{Phil. Mag. Ser. 6}  \textbf{27}, 315 (1914).
\newblock \doi{10.1080/14786440208635093}.

\bibitem{Darwin-PhilMag14b}
C.~Darwin.
\newblock \textrm{{LXXVIII.} the theory of {X-ray} reflexion. {Part II}}.
\newblock \emph{Phil. Mag. Ser. 6}  \textbf{27}, 675 (1914).
\newblock \doi{10.1080/14786440408635139}.

\bibitem{Ewald-adp16.1}
P.~P. Ewald.
\newblock \textrm{{Zur Begr\"undung der Kristalloptik}}.
\newblock \emph{Ann. Phys.-Leipzig}  \textbf{354}, 1 (1916).
\newblock \doi{10.1002/andp.19163540102}.
\newblock IV Folge Band 49; Einleitung zu Teil I (Dispersionstheorie) und Teil
  II (Theorie der Reflexion und Brechung).

\bibitem{Ewald-adp16.2}
P.~P. Ewald.
\newblock \textrm{{Zur Begr\"undung der Kristalloptik}}.
\newblock \emph{Ann. Phys.-Leipzig}  \textbf{354}, 117 (1916).
\newblock \doi{10.1002/andp.19163540202}.
\newblock IV Folge Band 49; Teil II:Theorie der Reflexion und Brechung.

\bibitem{Ewald-adp17.1}
P.~P. Ewald.
\newblock \textrm{{Zur Begr\"undung der Kristalloptik}}.
\newblock \emph{Ann. Phys.-Leipzig}  \textbf{359}, 557 (1917).
\newblock \doi{10.1002/andp.19173592402}.
\newblock IV Folge Band 54; Teil III: Die Kristalloptik der R\"ontgenstrahlen
  (Fortsetzung).

\bibitem{Ewald-adp17.2}
P.~P. Ewald.
\newblock \textrm{{Zur Begr\"undung der Kristalloptik}}.
\newblock \emph{Ann. Phys.-Leipzig}  \textbf{359}, 519 (1917).
\newblock \doi{10.1002/andp.19173592305}.
\newblock IV Folge Band 54; Teil III: Die Kristalloptik der R\"ontgenstrahlen.

\bibitem{Bethe-adp28}
H.~Bethe.
\newblock \textrm{{Theorie der Beugung von Elektronen an Kristallen}}.
\newblock \emph{Ann. Phys.-Leipzig}  \textbf{392}, 55 (1928).
\newblock \doi{10.1002/andp.19283921704}.

\bibitem{Bloch-zfp28}
F.~Bloch.
\newblock \textrm{{\"Uber die Quantenmechanik der Elektronen in
  Kristallgittern}}.
\newblock \emph{Zeitschrift f\"ur Physik}  \textbf{52}, 555 (1928).
\newblock \doi{10.1007/BF01339455}.

\bibitem{Sheridan-Optik90b}
J.~Sheridan and C.~Sheppard.
\newblock \textrm{An examination of the theories for the calculation of
  diffraction by square-wave gratings. {1. Thickness and Period Variations for
  Normal Incidence}}.
\newblock \emph{Optik}  \textbf{85}, 25 (1990).

\bibitem{Sheridan-Optik90a}
J.~Sheridan and C.~Sheppard.
\newblock \textrm{An examination of the theories for the calculation of
  diffraction by square-wave gratings. {2. Angular Variation}}.
\newblock \emph{Optik}  \textbf{85}, 57 (1990).

\bibitem{Sheridan-Optik90}
J.~Sheridan and C.~Sheppard.
\newblock \textrm{An examination of the theories for the calculation of
  diffraction by square-wave gratings. {3. Approximate Theories}}.
\newblock \emph{Optik}  \textbf{85}, 135 (1990).

\bibitem{Sheridan-josaa92}
J.~T. Sheridan and L.~Solymar.
\newblock \textrm{Diffraction by volume gratings - approximate solution in terms
  of boundary diffraction coefficients}.
\newblock \emph{J. Opt. Soc. Am. A}  \textbf{9}, 1586 (1992).
\newblock \doi{10.1364/JOSAA.9.001586}.

\bibitem{Sheridan-oc92}
J.~Sheridan and L.~Solymar.
\newblock \textrm{Spurious beams in dielectric gratings of the reflection type -
  a solution in terms of boundary diffraction coefficients}.
\newblock \emph{Opt. Commun.}  \textbf{94}, 8 (1992).
\newblock \doi{10.1016/0030-4018(92)90396-9}.

\bibitem{Sheridan-josaa93}
J.~Sheridan and C.~Sheppard.
\newblock \textrm{Coherent imaging of periodic thick fine isolated structures}.
\newblock \emph{J. Opt. Soc. Am. A}  \textbf{10}, 614 (1993).
\newblock \doi{10.1364/JOSAA.10.000614}.

\bibitem{Sheridan-Optik93}
J.~Sheridan.
\newblock \textrm{Stacked volume holographic gratings. {1. Transmission gratings
  in series}}.
\newblock \emph{Optik}  \textbf{95}, 73 (1993).

\bibitem{Sheridan-Optik94}
J.~Sheridan.
\newblock \textrm{Stacked volume holographic gratings. {2. Reflection gratings in
  series}}.
\newblock \emph{Optik}  \textbf{96}, 1 (1994).

\bibitem{Sheridan-oc94}
J.~Sheridan and C.~Sheppard.
\newblock \textrm{Modeling of images of square-wave gratings and isolated edges
  using rigorous diffraction theory}.
\newblock \emph{Opt. Commun.}  \textbf{105}, 367 (1994).
\newblock \doi{10.1016/0030-4018(94)90411-1}.

\bibitem{Sheridan-josaa94}
J.~Sheridan.
\newblock \textrm{Generalization of the boundary diffraction method for volume
  gratings}.
\newblock \emph{J. Opt. Soc. Am. A}  \textbf{11}, 649 (1994).
\newblock \doi{10.1364/JOSAA.11.000649}.

\bibitem{Raman-piasa36}
C.~V. Raman and N.~S.~N. Nath.
\newblock \textrm{The diffraction of light by high frequency sound waves: {Part
  I}}.
\newblock \emph{Proc. Ind. Acad. Sci. (A)}  \textbf{A2}, 406 (1936).
\newblock \doi{10.1007/BF03035840}.

\bibitem{Raman.2-piasa36}
C.~V. Raman and N.~S.~N. Nath.
\newblock \textrm{The diffraction of light by sound waves of high frequency:
  {Part II}}.
\newblock \emph{Proc. Ind. Acad. Sci. (A)}  \textbf{A2}, 413 (1936).
\newblock \doi{10.1007/BF03035841}.

\bibitem{Goodman-05}
J.~W. Goodman.
\newblock \textrm{Introduction to Fourier Optics}.
\newblock Roberts \& Company, Englewood, Colorado (2005).

\bibitem{Zachariasen-45}
W.~H. Zachariasen.
\newblock \textrm{Theory of X-Ray diffraction in Crystals}.
\newblock John Wiley \& Sons (1945).

\bibitem{Batterman-rmp64}
B.~W. Batterman and H.~Cole.
\newblock \textrm{Dynamical diffraction of x rays by perfect crystals}.
\newblock \emph{Rev. Mod. Phys.}  \textbf{36}, 681 (1964).
\newblock \doi{10.1103/RevModPhys.36.681}.

\bibitem{Leith-josa62}
E.~N. Leith and J.~Upatnieks.
\newblock \textrm{Reconstructed wavefronts and communication theory}.
\newblock \emph{J. Opt. Soc. Am.}  \textbf{52}, 1123 (1962).
\newblock \doi{10.1364/JOSA.52.001123}.

\bibitem{Chen-apl68}
F.~S. Chen, J.~T. {la Macchia}, and D.~B. Fraser.
\newblock \textrm{Holographic storage in lithium niobate}.
\newblock \emph{Appl. Phys. Lett.}  \textbf{13}, 223 (1968).

\bibitem{Klein-jasa65}
W.~R. Klein, C.~B. Tipnis, and E.~A. Hiedemann.
\newblock \textrm{{Experimental Study of Fraunhofer Light Diffraction by
  Ultrasonic Beams of Moderately High Frequency at Oblique Incidence}}.
\newblock \emph{J. Acoust. Soc. Am.}  \textbf{38}, 229 (1965).
\newblock \doi{10.1121/1.1909641}.

\bibitem{Burckhardt-josa66}
C.~B. Burckhardt.
\newblock \textrm{Diffraction of a plane wave at a sinusoidally stratified
  dielectric grating}.
\newblock \emph{J. Opt. Soc. Am.}  \textbf{56}, 1502 (1966).
\newblock \doi{10.1364/JOSA.56.001502}.

\bibitem{Burckhardt-josa67}
C.~B. Burckhardt.
\newblock \textrm{Efficiency of a dielectric grating}.
\newblock \emph{J. Opt. Soc. Am.}  \textbf{57}, 601 (1967).
\newblock \doi{10.1364/JOSA.57.000601}.

\bibitem{Gabor-prsa68}
D.~Gabor and G.~W. Stroke.
\newblock \textrm{The theory of deep holograms}.
\newblock \emph{Proc. Roy. Soc. A}  \textbf{304}, 275–89 (1968).
\newblock \doi{10.1098/rspa.1968.0086}.

\bibitem{Phariseau-piasa56}
P.~Phariseau.
\newblock \textrm{On the diffraction of light by progressive supersonic waves}.
\newblock \emph{Proc. Ind. Acad. Sci. (A)}  \textbf{44}, 165 (1956).

\bibitem{Syms-ao83}
R.~R.~A. Syms and L.~Solymar.
\newblock \textrm{Planar volume phase holograms formed in bleached photographic
  emulsions}.
\newblock \emph{Appl. Optics}  \textbf{22}, 1479 (1983).
\newblock \doi{10.1364/AO.22.001479}.

\bibitem{Syms-oa85}
R.~R.~A. Syms.
\newblock \textrm{Vector effects in holographic optical elements}.
\newblock \emph{Opt. Acta}  \textbf{32}, 1413 (1985).
\newblock \doi{10.1080/713821663}.

\bibitem{Syms-90}
R.~R.~A. Syms.
\newblock \textrm{Practical Volume Holography}.
\newblock Oxford University Press, Oxford (1990).

\bibitem{Moharam-josa82}
M.~G. Moharam and T.~K. Gaylord.
\newblock \textrm{Chain-matrix analysis of arbitrary-thickness dielectric
  reflection gratings}.
\newblock \emph{J. Opt. Soc. Am.}  \textbf{72}, 187 (1982).
\newblock \doi{10.1364/JOSA.72.000187}.

\bibitem{Moharam-josa82a}
M.~G. Moharam and T.~K. Gaylord.
\newblock \textrm{Diffraction analysis of dielectric surface-relief gratings}.
\newblock \emph{J. Opt. Soc. Am.}  \textbf{72}, 1385 (1982).
\newblock \doi{10.1364/JOSA.72.001385}.

\bibitem{Moharam-josaa86}
M.~G. Moharam and T.~K. Gaylord.
\newblock \textrm{Rigorous coupled-wave analysis of metallic surface-relief
  gratings}.
\newblock \emph{J. Opt. Soc. Am. A}  \textbf{3}, 1780 (1986).
\newblock \doi{10.1364/JOSAA.3.001780}.

\bibitem{Moharam-josaa95.02}
M.~G. Moharam, D.~A. Pommet, E.~B. Grann, and T.~K. Gaylord.
\newblock \textrm{Stable implementation of the rigorous coupled-wave analysis for
  surface-relief gratings - enhanced transmittance matrix approach}.
\newblock \emph{J. Opt. Soc. Am. A}  \textbf{12}, 1077 (1995).
\newblock \doi{10.1364/JOSAA.12.001077}.

\bibitem{Lalanne-josaa96}
P.~Lalanne and G.~M. Morris.
\newblock \textrm{Highly improved convergence of the coupled-wave method for {TM}
  polarization}.
\newblock \emph{J. Opt. Soc. Am. A}  \textbf{13}, 779 (1996).
\newblock \doi{10.1364/JOSAA.13.000779}.

\bibitem{Gaylord-ao81}
T.~K. Gaylord and M.~G. Moharam.
\newblock \textrm{Thin and thick gratings: terminology clarification}.
\newblock \emph{Appl. Optics}  \textbf{20}, 3271 (1981).
\newblock \doi{10.1364/AO.20.003271}.

\bibitem{Moharam_b-oc80}
M.~G. Moharam, T.~K. Gaylord, and R.~Magnusson.
\newblock \textrm{Criteria for {Bragg} regime diffraction by phase gratings}.
\newblock \emph{Opt. Commun.}  \textbf{32}, 14 (1980).
\newblock \doi{10.1016/0030-4018(80)90304-1}.

\bibitem{Moharam_rn-oc80}
M.~G. Moharam, T.~K. Gaylord, and R.~Magnusson.
\newblock \textrm{Criteria for {Raman-Nath} regime diffraction by phase
  gratings}.
\newblock \emph{Opt. Commun.}  \textbf{32}, 19 (1980).
\newblock \doi{10.1016/0030-4018(80)90305-3}.

\bibitem{Russell-ap80}
P.~S.~J. Russell and L.~Solymar.
\newblock \textrm{{Borrmann}-like anomalous effects in volume holography}.
\newblock \emph{Appl. Phys.}  \textbf{22}, 335 (1980).

\bibitem{Montemezzani-pre97}
G.~Montemezzani and M.~Zgonik.
\newblock \textrm{Light diffraction at mixed phase and absorption gratings in
  anisotropic media for arbitrary geometries}.
\newblock \emph{Phys. Rev. E}  \textbf{55}, 1035 (1997).
\newblock \doi{10.1103/PhysRevE.55.1035}.

\bibitem{Neipp-jpd02}
C.~Neipp, C.~Pascual, and A.~Bel{\'e}ndez.
\newblock \textrm{Mixed phase-amplitude holographic gratings recorded in bleached
  silver halide materials}.
\newblock \emph{J. Phys. D Appl. Phys.}  \textbf{35}, 957 (2002).
\newblock \doi{10.1088/0022-3727/35/10/303}.

\bibitem{Neipp-oex02}
C.~Neipp, I.~Pascual, and A.~Bel{\'e}ndez.
\newblock \textrm{Experimental evidence of mixed gratings with a phase difference
  between the phase and amplitude grating in volume holograms}.
\newblock \emph{Opt. Express}  \textbf{10}, 1374 (2002).
\newblock \doi{10.1364/OE.10.001374}.

\bibitem{Ellabban-spie07}
M.~A. Ellabban, M.~Bichler, M.~Fally, and I.~Dreven\v{s}ek~Olenik.
\newblock \textrm{Role of optical extinction in holographic polymer-dispersed
  liquid crystals}.
\newblock In: M.~Glogarova, P.~Palffy-Muhoray, and M.~Copic, eds., \emph{Liquid
  Crystals and Applications in Optics}, vol. 6587, 65871J:1--65871J:8. SPIE
  Proc. (2007).
\newblock \doi{10.1117/12.723361}.

\bibitem{Flauger-polymers19}
P.~Flauger, M.~A. Ellabban, G.~Glavan, J.~Klepp, C.~Pruner, T.~Jenke,
  P.~Geltenbort, and M.~Fally.
\newblock \textrm{Light- and neutron-optical properties of holographic
  transmission gratings from polymer-ionic liquid composites with submicron
  grating spacing}.
\newblock \emph{Polymers}  \textbf{11}, 1459 (2019).
\newblock \doi{10.3390/polym11091459}.

\bibitem{Neipp-jmo04}
C.~Neipp, M.~L. Alvarez, S.~Gallego, M.~Ortu\~no, J.~D. Sheridan, I.~Pascual,
  and A.~Beléndez.
\newblock \textrm{Angular responses of the first diffracted order in
  over-modulated volume diffraction gratings}.
\newblock \emph{J. Mod. Optic}  \textbf{51}, 1149 (2004).
\newblock \doi{10.1080/09500340408230413}.

\bibitem{Neipp-joa01}
C.~Neipp, I.~Pascual, and A.~Bel{\'e}ndez.
\newblock \textrm{Theoretical and experimental analysis of overmodulation effects
  in volume holograms recorded on {BB}-640 emulsions}.
\newblock \emph{J. Opt. A-Pure Appl. Op.}  \textbf{3}, 504 (2001).
\newblock \doi{10.1088/1464-4258/3/6/313}.

\bibitem{Neipp-ao01}
C.~Neipp, I.~Pascual, and A.~Bel\'endez.
\newblock \textrm{Effects of overmodulation in fixation-free rehalogenating
  bleached holograms}.
\newblock \emph{Appl. Optics}  \textbf{40}, 3402  (2001).
\newblock \doi{10.1364/AO.40.003402}.

\bibitem{Gallego-oc03}
S.~Gallego, M.~Ortu\~no, C.~Neipp, C.~Garc\'ia, A.~Bel\'endez, and I.~Pascual.
\newblock \textrm{Overmodulation effects in volume holograms recorded on
  photopolymers}.
\newblock \emph{Opt. Commun.}  \textbf{215}, 263 (2003).
\newblock \doi{10.1016/S0030-4018(02)02244-7}.

\bibitem{Neipp-Optik03}
C.~Neipp, M.~Alvarez, S.~Gallego, M.~Ortuno, I.~Pascual, and A.~Bel\'endez.
\newblock \textrm{Comparison between a thin matrix decomposition method and the
  rigorous coupled wave theory applied to volume diffraction gratings}.
\newblock \emph{Optik}  \textbf{114}, 529  (2003).
\newblock \doi{10.1078/0030-4026-00306}.

\bibitem{Neipp-oex03}
C.~Neipp, A.~Bel\'endez, S.~Gallego, M.~Ortu\~nuo, I.~Pascual, and J.~Sheridan.
\newblock \textrm{Angular responses of the first and second diffracted orders in
  transmission diffraction grating recorded on photopolymer material}.
\newblock \emph{Opt. Express}  \textbf{11}, 1835  (2003).
\newblock \doi{10.1364/OE.11.001835}.

\bibitem{Ortuno-ao03}
M.~Ortu\~no, S.~Gallego, C.~Gar\'cia, C.~Neipp, and I.~Pascual.
\newblock \textrm{Holographic characteristics of a 1-mm-thick photopolymer to be
  used in holographic memories}.
\newblock \emph{Appl. Optics}  \textbf{42}, 7008  (2003).
\newblock \doi{10.1364/AO.42.007008}.

\bibitem{Neipp-oc04}
C.~Neipp, J.~T. Sheridan, S.~Gallego, M.~Ortu\~no, A.~M\'arquez, I.~Pascual,
  and A.~Bel\'endez.
\newblock \textrm{Effect of a depth attenuated refractive index profile in the
  angular responses of the efficiency of higher orders in volume gratings
  recorded in a {PVA/acrylamide} photopolymer}.
\newblock \emph{Opt. Commun.}  \textbf{233}, 311 (2004).
\newblock \doi{10.1016/j.optcom.2004.01.064}.

\bibitem{Neipp-Optik04}
C.~Neipp.
\newblock \textrm{Thin and thick diffraction gratings: Thin matrix decomposition
  method}.
\newblock \emph{Optik}  \textbf{115}, 385  (2004).
\newblock \doi{10.1078/0030-4026-00382}.

\bibitem{Gallego-oex05}
S.~Gallego, M.~Ortu\~no, C.~Neipp, A.~M\'arquez, A.~Bel\'endez, I.~Pascual,
  J.~Kelly, and J.~Sheridan.
\newblock \textrm{Physical and effective optical thickness of holographic
  diffraction gratings recorded in photopolymers}.
\newblock \emph{Opt. Express}  \textbf{13}, 1939  (2005).
\newblock \doi{10.1364/OPEX.13.001939}.

\bibitem{Hernandez-oc06}
A.~Hern\'andez, C.~Neipp, A.~M\'arquez, S.~Gallego, I.~Pascual, and
  A.~Bel\'endez.
\newblock \textrm{Grating matrix method to describe a volume transmission
  diffraction grating}.
\newblock \emph{Opt. Commun.}  \textbf{266}, 122  (2006).
\newblock \doi{10.1016/j.optcom.2006.04.052}.

\bibitem{Gallego-Materials12}
S.~Gallego, C.~Neipp, L.~A. Estepa, M.~Ortu\~no, A.~M\'arquez, J.~Franc\'es,
  I.~Pascual, and A.~Bel\'endez.
\newblock \textrm{Volume holograms in photopolymers: Comparison between
  analytical and rigorous theories}.
\newblock \emph{Materials}  \textbf{5}, 1373 (2012).
\newblock \doi{10.3390/ma5081373}.
\end{thebibliography}
\end{document}